A Refined Model for Epitaxial Tilt of Epilayers Grown on Miscut Substrates

Michael E. Liao and Mark S. Goorsky

*University of California, Los Angeles, Los Angeles, California 90095, USA*

Abstract

A refined model of the origin of epitaxial tilt on miscut (or vicinal) substrates is explained by employing crystal modeling and reciprocal space analysis. The Nagai tilt model (H. Nagai, J. Appl. Phys., **45**, 3789 (1974)) is often cited to explain the tilt of lattice planes in a pseudomorphic layer deposited on a miscut substrate that is observed in high resolution x-ray diffraction measurements. Here, however, we demonstrate how that model incorrectly describes how the substrate applies biaxial stress onto the epitaxial layer. Most importantly, the stress applied to an epitaxial layer on a miscut substrate is not along a low index plane. For example, the surface plane of a nominally (001) cubic substrate with a miscut of 10° towards the [110] is the (118) plane and the stress applied is parallel along the (118) plane and not (001). Furthermore, under the framework of reciprocal space, the {00$l$} reflections would be symmetric reflections for on-axis substrates but asymmetric reflections for miscut substrates. The tilt that is experimentally observed between the low index substrate planes and the epitaxial layer planes ((001) for example with a miscut substrate) matches that which is predicted by treating the low index reflections as asymmetric reflections. An epitaxial tilt equation is provided which describes the tilt between epitaxial and substrate layers based on the lattice parameter mismatch as well as the Poisson ratio of the layer that is applicable to any crystal system.

Introduction

The Nagai tilt model[1] attempted to describe the origin of tilt in epitaxial layers grown on miscut (or vicinal) substrates. The basis of this model is the combination of atomic steps on a miscut substrate surface and the lattice parameter mismatch results in a tilted epitaxial layer planes with respect to the same planes in the substrate. While atomic steps are clearly observed on the surfaces of miscut substrates, we aim to demonstrate that that localized distortions at the step edge are not the origin of epitaxial tilt. The original model for epitaxial tilt by Nagai is reproduced in Figure 1(a), which suggests that the miscut substrate surface is along the treads of the atomic steps. According to this model, the directions of the biaxial stress that the substrate applies to the epitaxial layer would then be oriented perpendicular to these step ledges, i.e. parallel to the dashed lines as shown in Figure 1(a). For example, if Figure 1(a) represents a miscut (001) substrate, then the Nagai model indicates that the biaxial stress is applied to the epilayer along the in-plane directions of the (001) plane. The 'Nagai equation' which describes the tilt observed between epitaxial layer and substrate lattice planes is:

$$\alpha = \tan^{-1}\left\{\frac{\Delta a}{l}\right\}$$

where $\alpha$ is the angle between the orientation of the low index plane in the epitaxial layer compared to the orientation of the low index plane in the substrate, $\Delta a$ is the out-of-plane lattice parameter mismatch, and $l$ is the average step length.

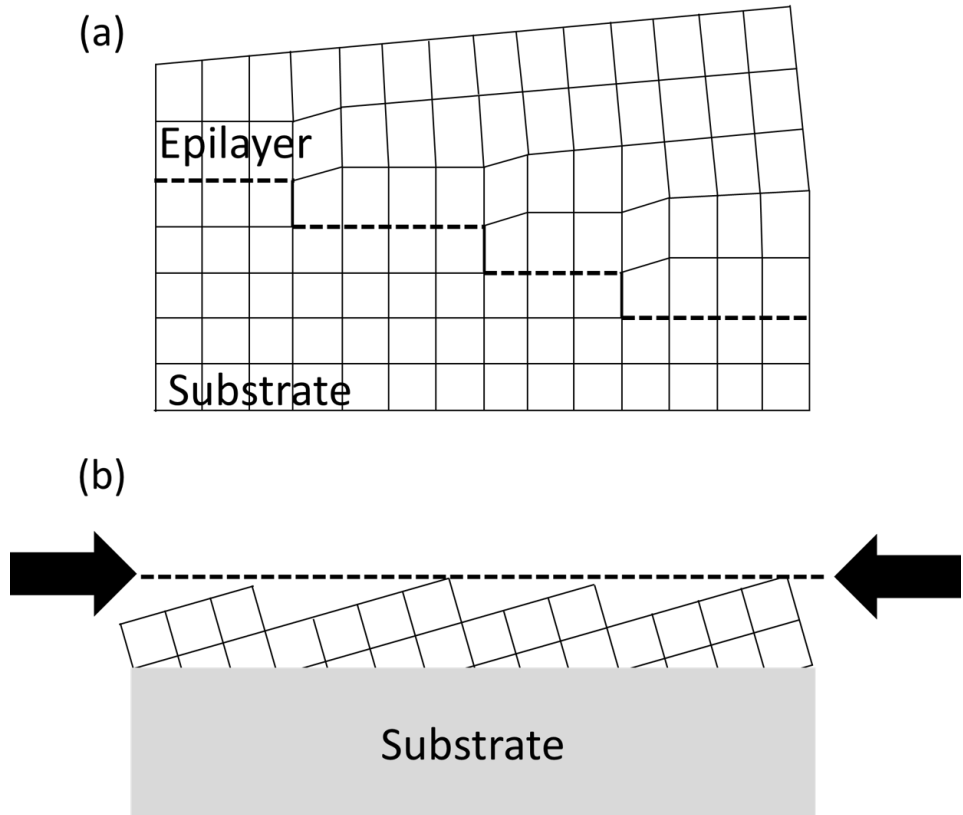

Figure 1. The Nagai tilt model (as presented by Nagai[1]) is shown in Fig. 1(a). The surface in this model is oriented along the treads of the atomic steps of the miscut substrate as indicated by the dashed lines, which shows that the applied biaxial stress imposed by the substrate is along the (001) plane. The actual substrate orientation of this miscut substrate is shown in (b) indicated by the dashed line. The black arrows represent the directions of the biaxial stress applied by the substrate.

However, the direction of the applied biaxial stress is parallel to the substrate surface for the general case where the front and back surfaces are parallel;[2,3] and for miscut or vicinal surfaces, for example, the stress is therefore not applied along the (001) plane. Figure 1(b) adjusts Nagai's original schematic diagram to show the true surface plane. In this example, the substrate is 10° miscut towards the [110] direction. An improved model representation of this 10° miscut substrate is shown in Figure 2, which uses silicon as an example.

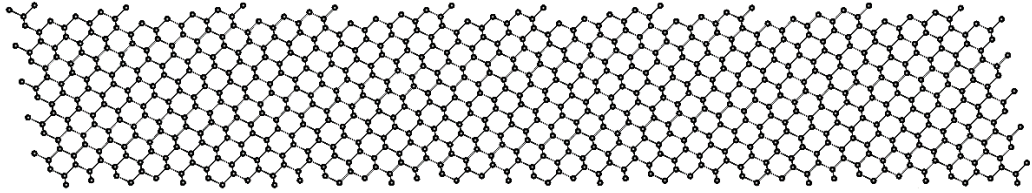

Figure 2. Crystal model for (001) Si miscut 10° towards the [110]. The true surface plane is (118).

The effective surface plane for this 10° miscut would then correspond to the (118) instead of (001) plane and stress induced by the lattice parameter difference between the substrate and epitaxial layer is parallel to this higher order plane ((118) here)., e.g. along the [44$\bar{1}$] and [1$\bar{1}$0] directions. Knowing the correct directions is crucial in understanding how the epilayer deforms in response to the biaxial stress from the substrate and why epitaxial tilt arises from this deformation. We also note that among various studies in the current literature that employ the Nagai tilt model, there are inconsistencies in the model diagrams of how the epitaxial lattice planes are deformed.[4,5,6] These differences are not trivial because differences in the structural aspects of how the epitaxial layer is deformed changes what one would expect to observe experimentally using X-ray diffraction (XRD) reciprocal space scans. For example, in Nagai's original tilt model shown in Figure 1(a), three peaks associated with the epitaxial layer would be expected: (1) a peak corresponding to the slight tilt that makes up most of the epitaxial layer, (2) a peak corresponding to the planes with no tilt parallel to the substrate steps, and (3) a peak corresponding to the relatively more severe tilt at the step corners (it should also be noted that these "severe tilt" regions reduce in magnitude with each subsequent layer, suggesting the presence of grading in this distortion along the surface normal). In the current literature, there have been no reports of observing multiple peaks associated with a single epitaxial layer due to the tilt variations suggested by the model depicted by Nagai. Thus, the Nagai model is a simple model but has not been observed experimentally.

Another significant contribution to epitaxial deformation relevant to epitaxial growth on miscut substrates is shear stress, which has been already studied and reported in the current literature.[7,8,9] De Caro et al.,[7] showed that substrates with surface planes that lack at least two-fold symmetry will induce shear stresses in epitaxial layers. Moreover, Lam et al.,[8,9] examined how shear in epitaxial layers manifests in reciprocal space for HgCdTe epilayers grown on (211) CdZnTe substrates. While the work presented in this paper focuses on pure tilt, combining our findings with the work on shear by De Caro et al.,[7] and Lam et al.,[8,9] provides a refined description of epitaxial deformation that can be applied to any arbitrary substrate orientation, crystal structure, and for various strain states ranging from fully strained (psuedomorphic) to fully relaxed.

Materials and Methods

A straightforward system will be used here while acknowledging that the choice of a crystal system can be arbitrary. The model substrate used is a (001) diamond cubic with a miscut

10° towards the [110]. An epitaxial layer with a ~20% difference in lattice parameter is also used. Systems with these properties would include silicon as the substrate and α-Sn as the layer. The high miscut and large mismatch are used here simply for a more clear visual demonstration of the effect. As shown later, the approach is equally valid for more realistic conditions – low miscut and smaller mismatch. Crystallographic models of the epitaxial layer were generated using Visualization for Electronic and STructural Analysis (VESTA).[10] The Sn layer was then fully strained to the miscut Si substrate. Reciprocal space representing both the on-axis and miscut Si substrate were also generated to show how a formerly symmetric reflection for an on-axis substrate becomes an asymmetric reflection for a miscut substrate. We then present a method for quantifying the amount of tilt expected in our Si-Sn example, which can be applied to other materials combinations for any crystal system.

Results

       The apparent tilt of an epitaxial layer can be measured using XRD by (1) first measuring the associated orientation plane with the substrate miscut direction parallel to the X-ray beam and then (2) measuring the same reflection after rotating the sample in-plane by 180° so the miscut direction is antiparallel to the incident X-ray beam. The angle the epitaxial layer peak displaces in these diffraction measurements is used to calculate tilt.[1] An important point is that this displacement in these measurements is also observed when performing measurements of asymmetric reflections (measurements performed in the glancing incident and glancing exit geometries). Glancing incident measurements of a given asymmetric reflection for an epitaxial layer appears farther away from the substrate peak than in glancing exit measurements along the diffraction scanning axis. This is due to the detector aperture oriented at a shallower angle with respect to the substrate surface compared to the detector aperture orientation in the glancing exit geometry. For a given asymmetric reflection, the glancing incident and exit geometries differ by an in-plane rotation of the sample by 180°.

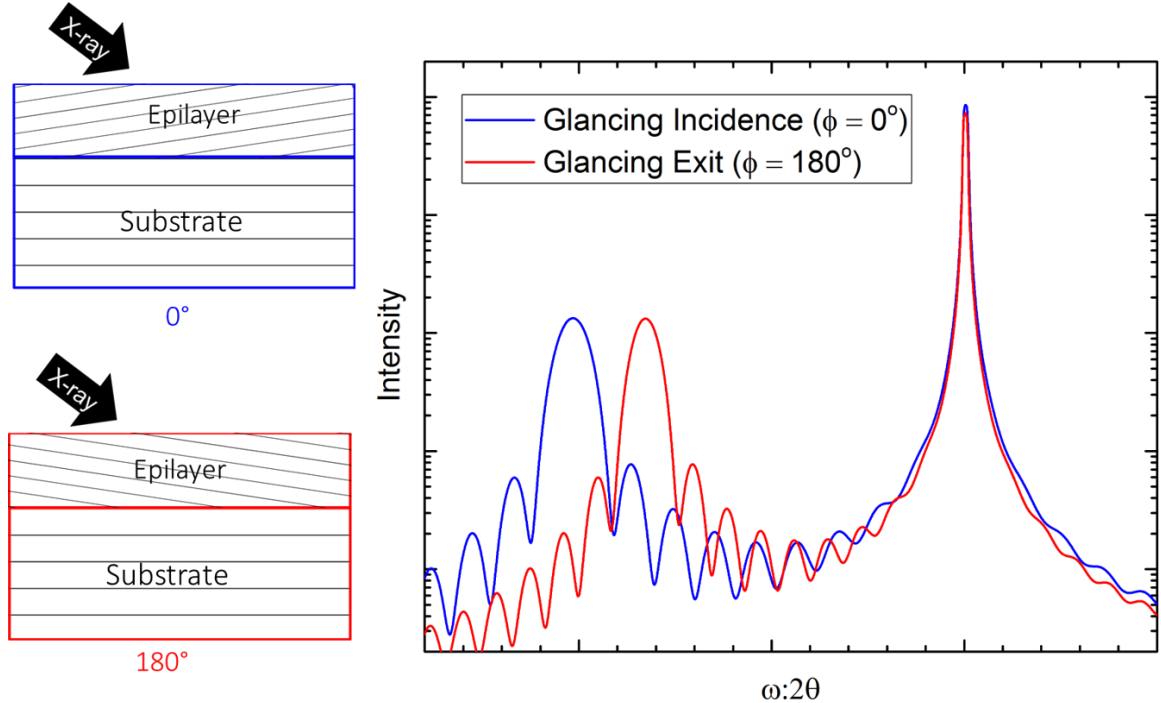

Figure 3. Diffraction simulation for the glancing incidence and glancing exit geometry. Note that for a given asymmetric reflection, the glancing incidence and glancing exit configurations differ by an in-plane rotation of 180°.

Under the framework of reciprocal space, it can be realized that the associated orientation plane of a miscut substrate actually corresponds to an asymmetric reflection as illustrated in Figure 4. Figure 4 shows reciprocal lattice points of (001) Si that is (a) on-axis,(b) 10° miscut towards [110], and (c) 15.8° miscut towards [110]. Note that the vertical dashed line corresponds to the zone axis normal to the surface, i.e. planes that correspond to symmetric reflections. As shown in Figure 4(b) and 4(c), the (004) reflection is actually an asymmetric reflection for a miscut (001) substrate. Figure 4(c) makes it apparent that a (001) substrate 15.8° miscut towards [110] is describing a (115) oriented substrate. Thus, by extension, a (001) substrate miscut 10° towards [110] is describing a (118) oriented substrate as shown in Figure 4(b). Furthermore, familiar orientations such as (111) and (110) oriented substrates can be viewed as a (001) substrate miscut 54.7° and 90° towards the [110], respectively. For these familiar orientations, it is easily understood that the (004) reflection is an asymmetric reflection. It is also important to note that for symmetric reflections, the peak splitting between the substrate and epitaxial layer is invariant to rotating the substrate 0° vs 180° in-plane, i.e. for epitaxially layers grown on on-axis substrates with no miscut because all symmetric reflections lie along the axis of rotation perpendicular to the surface. The axis for in-plane rotation corresponds to the vertical axes in Figure 4.

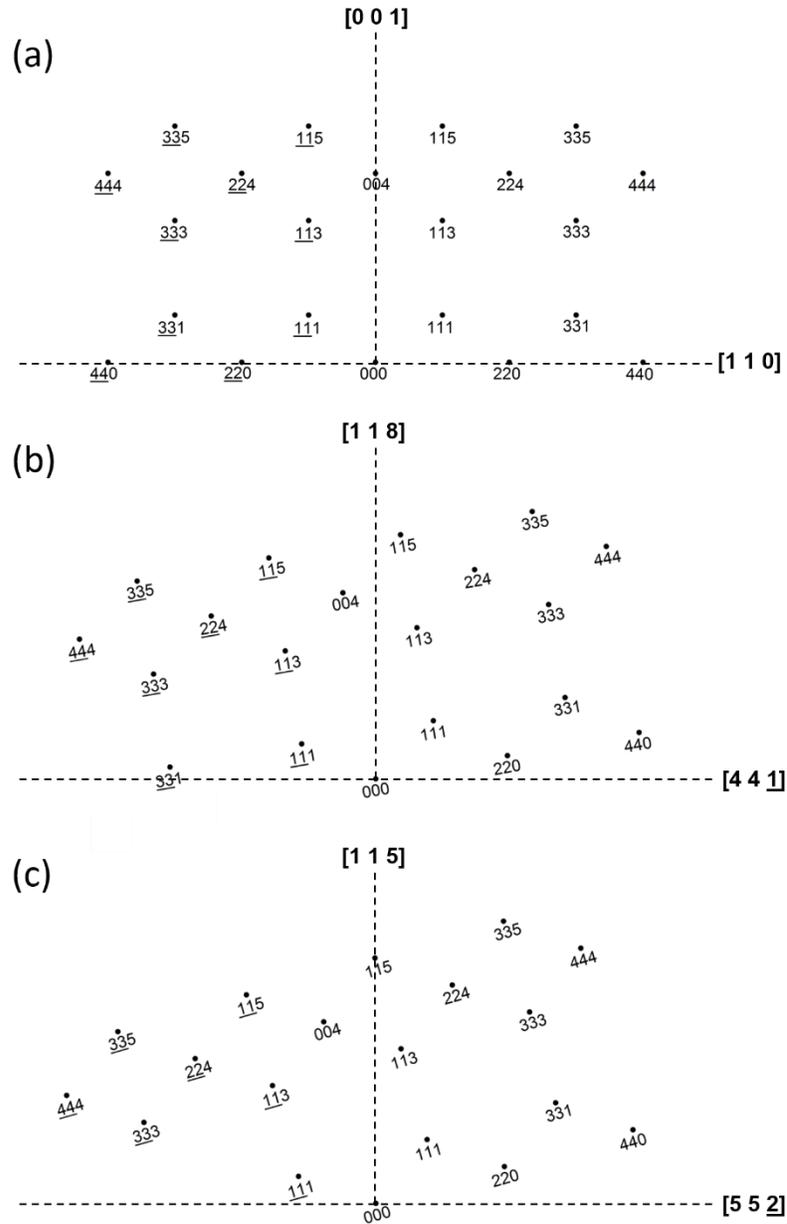

Figure 4. Reciprocal lattice points for (a) an on-axis (001) Si substrate, (b) a (001) Si substrate miscut 10° towards the [110], and (c) a (001) Si substrate miscut 15.8° towards the [110]. The horizontal dashed line corresponds to the direction parallel to the substrate surface while the vertical dashed line corresponds to the direction normal to the substrate surface.

These reciprocal space maps make it apparent that when biaxial stress is applied parallel to the substrate surface, the (004) reflection would not tilt for an on-axis substrate while tilt of the (004) is expected for the miscut substrate. This is due to the fact that in the case of the on-axis substrate, the {00*l*} planes only experience one component of lattice distortion (i.e. only along the surface normal) while for the miscut substrate, the {00*l*} planes experience two contributions of lattice distortion – (1) parallel to the surface plane and (2) perpendicular to the surface plane.

Next, a fully strained α-Sn epitaxial layer was simulated on a (001) Si substrate miscut 10° towards the [110] by building crystal models as shown in Figure 5. The true surface plane of this Si substrate is the (118) plane.

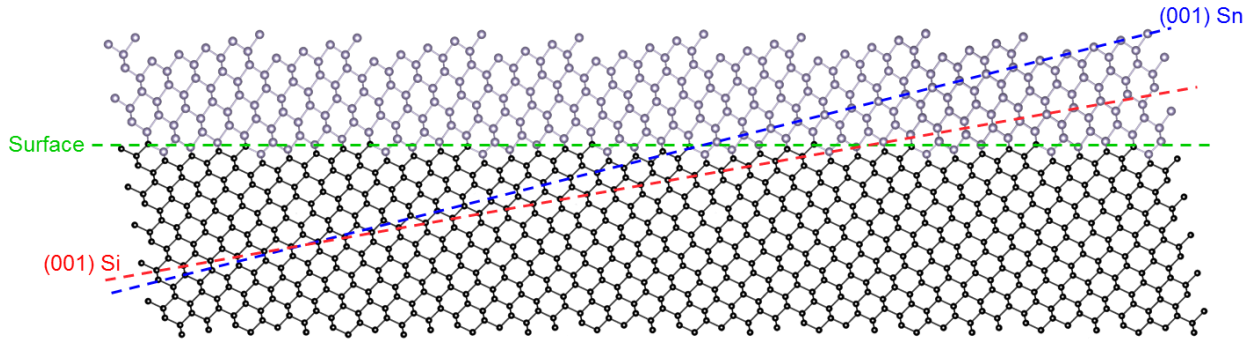

Figure 5. Crystal models of a fully strained Sn epitaxial layer on a (001) Si substrate miscut 10° towards the [110]. The black atoms are Si atoms while the gray atoms are Sn atoms. Note that the Sn atoms adjacent to the Si atoms at the heterointerface extend into the page while the Si atoms extend out of the page (tetrahedral sites in the diamond cubic structure).

As demonstrated with the reciprocal space maps in Figure 4, the crystal models also show how the {00*l*} planes tilt in response to both the biaxial strain applied in-plane and the resulting out-of-plane strain normal to the (118) plane In this case, the Sn {00*l*} planes tilt ~4°. The resulting change in the unit cell lattice parameters for the strained Sn due to the biaxial strain can be calculated. In this example, for a fully strained Sn epitaxial layer on this miscut (001) Si substrate, the lattice parameters of Sn change from: (relaxed parameters) $a = b = c = 0.6489$ nm and $\alpha = \beta = \gamma = 90°$ to (strained parameters) $a = b = 0.5472$ nm, $c = 0.7622$ nm, $\alpha = \beta = 85.1°$, and $\gamma = 89.1°$. The relaxed and fully strained Sn unit cells are shown in Figure 6. The angle α corresponds to the angle between the b- and c-axes, β corresponds to the angle between the a- and c-axes, and γ corresponds to the angle between the a- and b-axes.

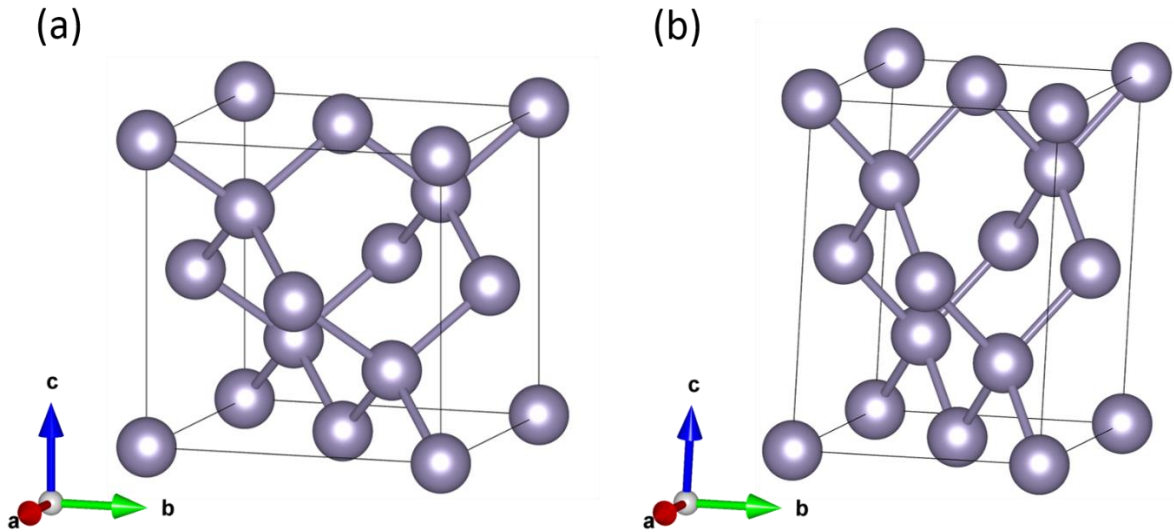

Figure 6. Unit cell of (a) relaxed α-Sn and (b) the unit cell after applying biaxial stress along the (118) plane fully strained on the 10° miscut Si substrate.

These crystal models also illustrate the orientations of the planes as well as the relationship between the Sn atoms and Si atoms especially across the interface. Figure 7 is a magnified version of Figure 5 focused on the Sn||Si interface.

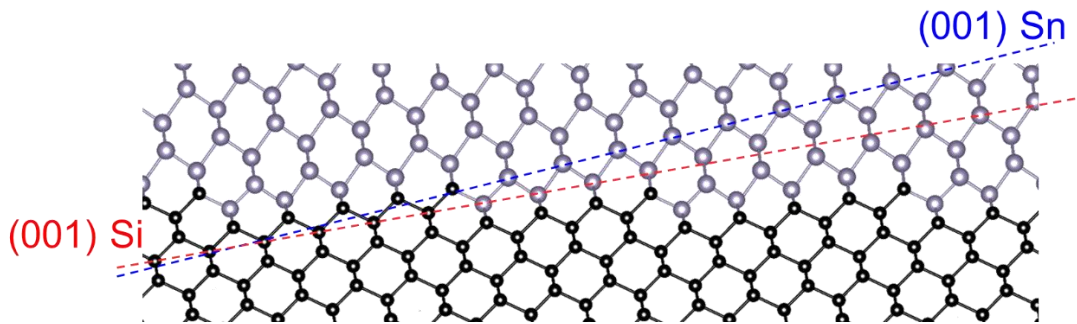

Figure 7. The fully strained Sn epitaxial layer on Si shown in Figure 5 magnified at the heterointerface. Dotted lines have been added to guide the eye in following the (001) Si planes and (001) Sn planes.

Note that the Sn atoms adjacent to the Si atoms at the interface are spatially oriented into the page while the Si atoms are oriented out of the page (i.e. the tetrahedral sites in the diamond cubic structure). While the Si-Sn system may be an experimentally extreme example, the large lattice mismatch and miscut angle help visually showcase the structural distortion of the atomic bonds (at the interface and in the bulk of the Sn epitaxial layer) and consequently the lattice planes. Furthermore, building crystal models has the advantage over the simplified Nagai diagrams in providing this detailed structural information and modeling of the atomic bonds and how all the lattice planes may align between the epitaxial layer and substrate. Details in modeling are important for calculating behavior characteristics heavily influenced by structural changes to the unit cell (i.e. strain) such as electronic[11,12] and thermal transport.[13,14] Additionally, this would especially be useful for modeling and understanding other complicated noncubic, lower-symmetry crystals.

Here we outline the method for calculating the expected tilt for any strain state of the epitaxial layer. (1) Identify the directions normal and parallel to the surface. For this example of the (001) Si substrate miscut 10° towards the [110], the directions would be [118] and [44$\underline{1}$], respectively. (2) Calculate the in-plane strain and out-of-plane strain for the epitaxial layer. (3) Calculate the components of these directions along the miscut direction, i.e. the [110] in this example. (4) Subtract the total epitaxial tilt with respect to its (118) surface from the substrate miscut angle to determine the amount of epitaxial tilt with respect to the substrate.

$$t = \tan^{-1}\left\{\frac{d_{e,o}(\vec{a}\cdot\vec{b})[d_{e,i}(1+\nu) - 2\nu d'_{e,i}]}{d_{e,i}d'_{e,i}(\vec{a}\cdot\vec{c})(1-\nu)}\right\} - m$$

where $t$ is the relative tilt of the epitaxial layer plane of interest; $d_{e,o}$ and $d_{e,i}$ are the relaxed out-of-plane and in-plane epitaxial layer d-spacing, respectively; $d'_{e,i}$ is the strained in-plane epitaxial layer d-spacing; $\nu$ is the epitaxial layer Poisson's ratio; $m$ is the miscut angle of the substrate; $\vec{a}$ is the direction parallel to the associated orientation plane along the miscut direction; $\vec{b}$ is the direction perpendicular to the substrate's true surface; and $\vec{c}$ is the direction parallel to the substrate's true surface along the miscut direction.

Next, the quantitative results of the tilt calculation method presented here with the equation associated with the Nagai model were compared for four pseudomorphic scenarios: (1) low lattice mismatch and 0.1° miscut, (2) low lattice mismatch and 10° miscut, (3) high lattice mismatch and 0.1° miscut, and (4) high lattice mismatch and 10° miscut. The low lattice mismatch corresponds to a lattice parameter difference of ~0.14% (e.g. AlAs on GaAs) while the high lattice mismatch corresponds to a difference of ~19.5% (e.g. α-Sn on Si). Tilt was calculated using both models for Poisson's ratios ranging from 0 to 0.5 as shown in Figure 8.

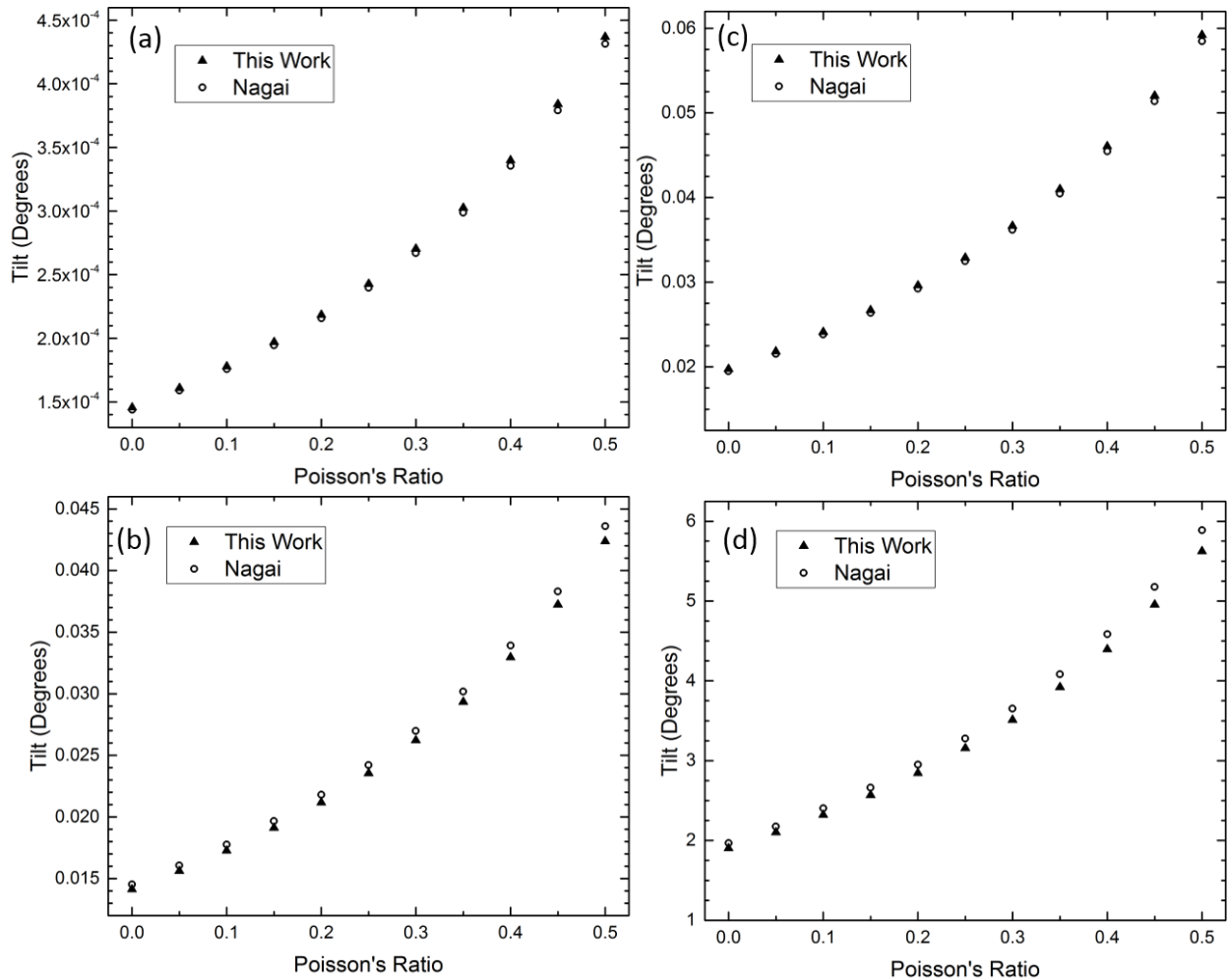

Figure 8. Comparison of the expected tilt for various lattice mismatch combinations and substrate miscut angles: (a) 0.14% lattice mismatch and 0.1° miscut, (b) 0.14% lattice mismatch and 10° miscut, (c) 19.5% lattice mismatch and 0.1° miscut, and (d) 19.5% lattice mismatch and 10° miscut.

For 8(a) the low lattice mismatch and 0.1° miscut case, the difference between our method and Nagai's equation is negligible at 1.3% – with the magnitude of that difference corresponding to 0.01" to 0.02". For 8(b) the low lattice mismatch and 10° miscut case, the difference is 2.9%, which corresponds to a magnitude of 1.5" to 4.4". On the other hand, 8(c) the high lattice mismatch and 0.1° miscut corresponds to a percent difference of 1.3% with a magnitude of 0.9" to 2.7". Lastly, for 8(d) the high lattice mismatch and 10° miscut case, the difference ranges from 3.5% to 4.7%, which corresponds to a magnitude difference of 0.07° to 0.26°. While the Nagai equation may give acceptable values especially for low lattice mismatch and low miscut cases, the picture it provides at the interface and in the bulk of the epitaxial layer is not accurate. The distortions around the atomic step ledges are not the cause of epitaxial tilt as the Nagai model suggests. Instead, it is the direction of the biaxial stress applied along the true surface plane, which is a high index plane for a miscut substrate. The low index plane, e.g. the (001) epitaxial layer planes on a miscut substrate, experiences both in-plane and out-of-plane distortion contributions which results in the tilt of the low index planes.

Conclusion

The model presented in this study provides a clearer picture of the growth interface to illustrate the origin of epitaxial tilt. The pseudomorphic case was examined in this study but can be extended to any strain state as well as for any crystal systems. Realizing the true surface plane enables us to identify the correct direction of the applied biaxial stress from the substrate that the Nagai tilt model fails to do. By doing so, further insight from reciprocal space enables us to understand that the low index plane associated with a miscut/vicinal substrate is an asymmetric reflection for miscut substrates – which then enables us to realize that this plane has both in-plane and out-of-plane components of distortion. While the Nagai equation may have been shown to yield acceptable numerical values for tilt in various cases reported in the literature, that model provides an inaccurate picture of the growth interface and how the substrate stresses the epitaxial layer. Having a precise picture is important for understanding interfaces especially because both interfaces and the structure of the lattice parameter govern key characteristics such as electronic and thermal transport across interfaces.

Acknowledgement

The authors would like to acknowledge the support from the Office of Naval Research through a MURI program, grant No. N00014-18-1-2429.

Data Availability

The data that support the findings of this study are available from the corresponding author upon reasonable request.